\documentclass[twocolumn,aps,prb,showpacs,amsmath]{revtex4}
\usepackage{amssymb}
\usepackage{mathrsfs}
\usepackage{graphicx}
\usepackage{float}

\begin{document}

\title{Topologically protected mid-gap states induced by impurity in one-dimensional superlattices}
\author{Li-Jun Lang}
\email{langlijun84@gmail.com}
\author{Shu Chen}
\email{schen@iphy.ac.cn}
\affiliation{Beijing National Laboratory for Condensed Matter Physics, Institute of
Physics, Chinese Academy of Sciences, Beijing 100190, China}
\begin{abstract}
Based on the discovery of the nontrivial topology of one-dimensional superlattices, we show that midgap states will emerge in such systems induced by a single on-site impurity. Besides the trivial bound state located at the impurity site, these midgap states are localized at the adjacent sides of the impurity, carrying the information of the topology and behaving like the effect of the open boundary conditions in the limit of a large attractive or repulsive impurity potential. Using this feature, the impurity can be used to detect the topology of the superlattice system and to realize the adiabatic pumping between the opposite sides of the impurity in cold-atom experiments or in photonic crystals.
\end{abstract}
\date{\today}
\maketitle

\section{Introduction}
In recent years, due to the very fast development of the cold atom
techniques \cite{IBloch2008}, quantum simulation of topological
insulators in cold atomic systems has been attracting more and
more attentions experimentally and theoretically \cite{TI}.
However, while nearly all studies focus on the two-dimensional or
three-dimensional systems, one-dimensional (1D) topological
insulators are less known besides the Su-Schrieffer-Heeder model
\cite{Su1979}. Recently, the discovery of the topology of 1D
superlattices adds a new member \cite{Me2012, YEKraus2012}.
Because of the easy realization of the superlattices both in cold
atoms \cite{AndersonEx} and in photonic quasicrystals
\cite{YEKraus2012}, it paves the way of studying the topology of
1D superlattices, to date including topological phase transitions
\cite{TPT}, fractional topological states \cite{ZhihaoXu2013},
$Z_2$ topological insulators \cite{FengMei2012,SGaneshan},
topological Mott insulators \cite{TMI}, and even topological
superconductors \cite{Majorana}.

As we know that the detection of the topological edge states
depends on the open boundary conditions (OBCs) of the system
\cite{YHatsugai}, in cold atoms the confining harmonic trap blurs
the boundary effects, making the edge states not obvious
\cite{MBuchhold2012}. To demonstrate the nontrivial topology of
the 1D superlattices, a variety of proposals have been given, such
as detecting the plateaus in the density profile subjected to a
harmonic trap in cold atoms \cite{Me2012}, and the direct
observation of the localized edge states using waveguides
\cite{YEKraus2012}. In this paper, we find that a single on-site
impurity will induce midgap states for the superlattices and they
are localized next to the impurity site in the limit of the large
impurity potential, behaving like the edge states under OBCs.
Using this feature, we can easily detect these states in the
interior of the system instead of at the boundaries, and also
realize the transportation of midgap states between the opposite
sides of the impurity. By the technique of the spatially resolved
radio-frequency spectroscopy (SRRFS) \cite{YShin2007}, the local
density of states (LDOS) can be measured directly. So the impurity
supplies a more direct and realistic scheme to study the topology
of the 1D superlattices.

The paper is organized as follows: After a brief introduction, in
Sec. \ref{Z-type} we first introduce the midgap states induced by
a single on-site impurity in 1D spinless fermionic superlattices
and show the relationship of these states to the $Z$-type topology
possessed by such superlattices. In Sec. \ref{coldatom}, the
behavior of the midgap states can be used to show the topology of
the superlattices by the technique of the SRRFS in cold atoms to
detect the LDOS in the harmonic trap. Sec. \ref{Z2-type}
generalizes the discussion to the spin-1/2 case and the
relationship of these states to the $Z_2$-type topology.
Alternatively, Sec. \ref{pump} gives another scheme to demonstrate
the topology of superlattices through midgap states using photonic
crystals and realizes the adiabatic pumping between opposite sides
of the impurity. In Sec. \ref{sum}, we have a summary.

\section{$Z$-type topological insulators with a single on-site impurity} \label{Z-type}

Firstly, we concentrate on the 1D tight-biding superlattices for
spinless fermions with $L$ lattice sites, doped with a single
on-site impurity. The Hamiltonian is as follows:
\begin{equation}
H=-t\sum_{i}(\hat{c}_{i}^{\dag }\hat{c}_{i+1}+\text{H.c.})+\sum_{i}V_{i}\hat{%
n}_{i}+F\hat{n}_{0},  \label{spinlessHam}
\end{equation}%
where $\hat{c}_{i}^{\dag }\ (\hat{c}_{i})$ is the creation (annihilation)
operator at site-$i$ for spinless fermions, and $\hat{n}_{i}=\hat{c}%
_{i}^{\dag }\hat{c}_{i}$. $t$ is the hopping amplitude and set to be the
unit of energy. $V_{i}=V\cos (2\pi \alpha ia+\delta )$ is the on-site
modulated potential with $V$ being the strength, $1/\alpha $ the
periodicity, and $\delta $ the arbitrary phase shift. $F$ is the potential
of the impurity. Without loss of generality, we choose odd number of lattice sites and put the impurity at the center of the chain (the center site is labeled as the origin) in the following discussion. The lattice
spacing, $a$, is taken to be $1$, for convenience.

\begin{figure}[h]
\includegraphics[width=1\linewidth]{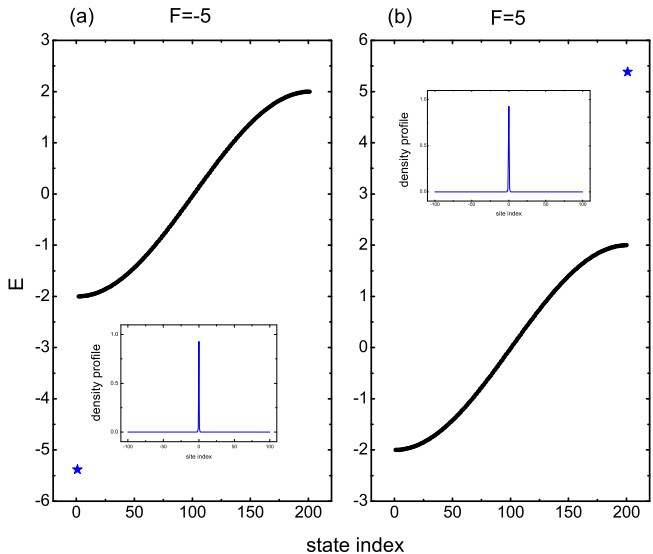}\newline
\caption{The single particle eigenenergies in ascending order with (a) the attractive ($F=-5$) and (b) the repulsive ($F=5$) impurities located at the center of the 1D tight-binding chain without modulation under PBCs. The insets are density profiles of the bound states labeled by blue stars, which are localized at the impurity site with energies (a) lower or (b) higher than that of the scattering continuous. Here we choose $t=1, L=201$.}
\label{nonmod}
\end{figure}

\begin{figure}[h]
\includegraphics[width=1\linewidth]{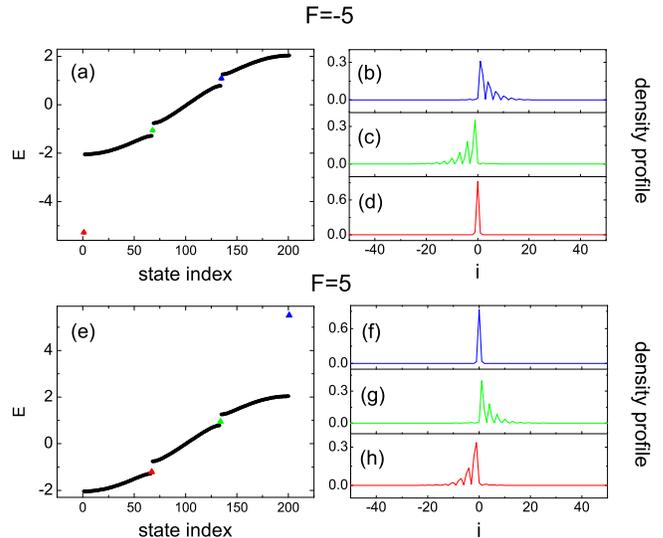}\newline
\caption{For the attractive ($F=-5$) (a-d) and the repulsive ($F=5$) (e-h) impurity with the periodic modulation, $\alpha=1/3$, the eigenenergies with the midgap states labeled by colored triangles are shown in ascending order. The corresponding density profiles (b-d, f-h) of the midgap states in the same color demonstrate the ways of localization at one side of the impurity or the impurity site. The other parameters we choose are $t=1, L=201, V=0.5, \delta=\pi/4$ under PBCs.}
\label{modnm}
\end{figure}

Without the modulation, there is only one bound state localized at the impurity site with its energy lower ($F<0$) (Fig. \ref{nonmod}(a)) or higher ($F<0$) (Fig. \ref{nonmod}(b)) than that of the scatting states. When the superlattice is turned on, besides the trivial bound states, there are also some states emerging within the bulk gaps. These midgap states are also localized. However, they are not localized at the impurity site but on the ones nearest to it (Fig. \ref{modnm}). Because the behaviors of the cases for $F>0$ and $F<0$ are similar except for the energies of the trivial bound states, we mainly focus on the $F<0$ case below.

We have known that without the impurity, this 1D superlattice
system is topologically non-trivial \cite{Me2012, YEKraus2012},
which is clearly seen when we adiabatically shift the phase,
$\delta$, under OBCs the edge states will connect the nearest bulk
bands and thus transfer from one end of the chain to the other. A
$Z$-type topological invariant, Chern number, can be defined in
the $(k, \delta)$-space under periodic boundary conditions (PBCs)
\cite{Me2012}, i.e.,
$C=\frac{1}{2\pi}\int_{0}^{{2\pi}/{q}}{dk}\int_{0}^{2\pi} {d
\delta} [\partial_k A_\delta -\partial_\delta A_k ]$ with the
Berry connection $A_s = i \langle \phi(k,\delta)|\partial_s
|\phi(k,\delta) \rangle $ $(s=\delta,k)$, where $k$ is the
quasimomentum and $\phi(k,\delta)$ is the Bloch wavefunction of
the Hamiltonian for each $\delta$.

\begin{figure}[h]
\includegraphics[width=1\linewidth]{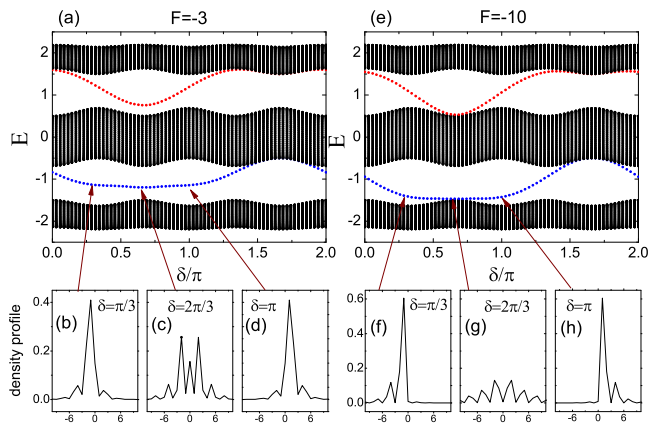}\newline
\caption{The eigenenergies versus $\delta$ for a single impurity
of a weaker ($F=-3$) (a-d) and a stronger ($F=-10$) (e-h) on-site
potentials under the periodic modulation, $\alpha=1/3$. The midgap
states are labeled by colored dots, which tend to connect the
nearest bulk bands and to localize at one side of the impurity in
the limit of a large impurity potential. The corresponding density
profiles show some specific point of $\delta$'s to demonstrate the
transfer from one side to the other. The other parameters we
choose are $t=1, L=201, V=1$ under PBCs.} \label{nmvd}
\end{figure}

Now if we turn on the impurity potential, the midgap states appear
in each bulk gap varying with the phase shift $\delta$ (Figs.
\ref{nmvd}(a) and \ref{nmvd}(e)), and as the strength of the
impurity potential, $V$, increases, these states will become more
and more localized at the adjacent sites of the impurity, and
connect the nearest bulk bands just like the edge states under
OBCs (Fig. \ref{nmvd}(e)). Different from the edge states, when we
shift the phase of the superlattice the midgap states do not
transfer from one edge to the other, but from one side of the
impurity to the other, which just happens at a very small space
helpful to the experimental detection. So naturally the single
on-site impurity is a good candidate for detecting the topological
properties of the superlattice systems, which has been proposed to
detect zero-mode Majorana fermions in $p$-wave topological
superconductors \cite{HuiHu2013,XiaJiLiu2013}.

\section{Detection of midgap states in cold atoms} \label{coldatom}

To realize the superlattice Hamiltonian (\ref{spinlessHam}) in
cold atoms  \cite{AndersonEx}, we first trap an atomic gas in a
primitive 1D optical lattice potential, $V_1 (x)=V_1 cos^2 (k_1
x)$, with the wave number $k_1$ and the strength $V_1$, then
superimpose a secondary optical lattice potential, $V(x)=Vcos(k_2
x+\delta)$, with wave number $k_2$ as a weak perterbation if
$V_1\gg V$. In the framework of the tight-binding approximation,
the primitive and the secondary optical lattices are mainly
supplying the hopping term and the on-site term of the Hamiltonian
(\ref{spinlessHam}), respectively. Here $\alpha=k_2/k_1$ can be
tuned by the wave numbers of the two sets of optical lattices and
$\delta$ is the overall relative phase shift between them
\cite{TSalger2007}.

We know that the observation of the edge states is difficult in
cold atom experiments due to the harmonic trap for confining
Bose-Einstein condensates (BECs), which weakens the boundary
conditions and reduces the particles at both wings of the trap.
Here we show that the impurity can induce midgap states and has
the similar effect of the OBCs in the uniform systems. These
midgap states hold the information of the topology of this system
and behave like the edge states only in a very small space as we
mentioned before and overcome the shortcomings from the harmonic
trap. So placing an impurity into the center of the harmonic trap
can transfer the detection of the topological edge states from
boundaries of the system to the interior, and the localization of
the midgap states will be much more notable than those when they
are at edges.

We calculate the LDOS in Lehmann representation at zero temperature,
\begin{equation}
\rho(i,\omega)=\sum_{n}[|\left\langle G|c_{i}|n\right\rangle |^{2}%
\delta(\omega-\omega_{n0})+|\left\langle G\right\vert c_{i}^{\dag}\left\vert
n\right\rangle |^{2}\delta(\omega+\omega_{n0})],\label{LDOS}%
\end{equation}
where $\omega _{n0}=E_{n}-E_{0}$ with $\left\vert G\right\rangle $ and $%
\left\vert n\right\rangle $ being the ground state with energy
$E_{0}$ and the $n$-th excited state with energy $E_{n}$,
respectively. This quantity can be directly measured
experimentally by SRRFS \cite{YShin2007}. Fig. \ref{LDOS} shows
the LDOS after adding the harmonic trap term, $H_h=V_h i^2$ with
$V_h$ being the strength of the trap. We can see that as we tune
the $\delta$, the midgap states are firstly localized at one side
of the impurity, then merge into the bulk bands becoming extended,
and finally transfer to the other side of the impurity.
\begin{figure}[h]
  \includegraphics[width=1\linewidth]{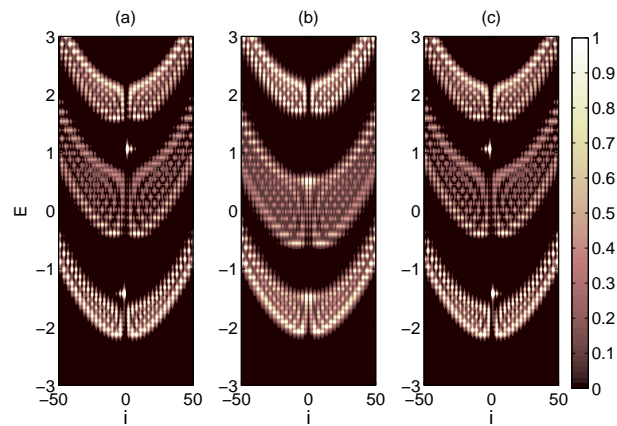}\\
  \caption{LDOS for spinless fermions with a single impurity of a large on-site potential ($F=-10$) within a slowly varying harmonic trap ($V_h=0.0005$). The midgap states in the bulk gaps are transferred from one side of the impurity to the other by tuning the phase shift, (a) $\delta=\pi/3$, (b) $\delta=2\pi/3$, (c) $\delta=\pi$. Here we take $\alpha=1/3, t=1, L=201, V=1$.}\label{LDOS}
\end{figure}

\section{$Z_2$-type topological insulators with a single on-site impurity} \label{Z2-type}

For spin-$1/2$ fermions, a spin-dependent secondary optical
lattice is added with the form $V(x)=Vcos(2\pi\alpha x
\pm\delta)$, which can be implemented by two counter-propagating
laser beams with linearly polarizations forming an angle $2\delta$
\cite{VFinkelstein1992, DJaksch1999}. So Hamiltonian
(\ref{spinlessHam}) can be extended to the spin-dependent one just
regarding $\hat{c}_{i}=(\hat{c}_{i\uparrow},
\hat{c}_{i\downarrow})^T$ as a two-component spinor for spin-up
and spin-down fermionic operators and $V_i$ as
$V_i=\text{diag}(Vcos(2\pi\alpha i+\delta),Vcos(2\pi\alpha
i-\delta))$. From the same procedure like the 1D spinless
superlattice to simulate the quantum Hall effect and demonstrate
the $Z$-type topology, this generalized 1D spin-dependent
superlattice can be used to simulate the quantum spin Hall effect
and demonstrate the $Z_2$-type topological insulators
\cite{FengMei2012}. Likewise, here the spin-dependent Chern number
can be also defined in the $(k, \delta)$-space under PBCs as
$C_\sigma=\frac{1}{2\pi}\int_{0}^{{2\pi}/{q}}{dk}\int_{0}^{2\pi}
{d \delta} [\partial_k A_{\delta\sigma} -\partial_\delta
A_{k\sigma} ]\ (\sigma=\uparrow \text{or} \downarrow)$ with the
spin-dependent Berry connection $A_{s\sigma}$ and the Bloch wave
function $\phi(k,\delta)$ for fixed $\delta$. Using these
spin-dependent Chern numbers, a total $Z_2$ topological invariant
is constructed as $\nu=(C_\uparrow-C_\downarrow)/2$, which
demonstrates the $Z_2$-type topology. Furthermore, this formalism
is also applicable to spin-orbit systems where the total spin does
not conserve, the only change of these definitions for the
topological invariants is to replace the fermions by quasi-femions
which are the mixture of the spin-up and spin-down fermions with
pseudo-spins. As long as the bulk gap does not close, the topology
of the system will not change.

When the impurity is embedded, because of the presence of the spin-dependent potentials of the superlattice, different from the spinless case, the system may possess simultaneously spin-up and spin-down fermions localized at one side or both sides of the impurity. In principle, we can always translate the whole superlattice by a phase, $\delta_0$, which makes the midstates shift wholly by $\delta_0$, i.e., $V_i=\text{diag}(Vcos(2\pi\alpha i-\delta_0+\delta),Vcos(2\pi\alpha i-\delta_0-\delta))$. By tuning the whole phase $\delta_0$ and the relative phase $\delta$, we can manipulate the emergence of the midgap states at one side or both sides of the impurity simultaneously or non-simultaneously. For example, without the phase difference of the two component fermions, the whole spectrum is of double degeneracy with spin-up and spin-down fermions having the same energy, and the midgap states will always be localized at the same side of the impurity for fixed $\delta_0$. Another case is that we can always make the whole spectrum mirror-symmetric with respect to $\delta=\pi$ by tuning $\delta_0$, e.g. $\delta_0=\pi/3$ for the case of $\alpha=1/3$. Thus the spectrum for fixed $\delta$ is still of double degeneracy with each spin of the midgap states not localized at the same side of the impurity but at the opposite side separatively and simultaneously. Fine-tuning the both phases, we may get a rich physics. Here we mainly demonstrate the physics of the second case with $\delta_0=\pi/3$ for the case of $\alpha=1/3$ to make the whole spectrum mirror-symmetric.

\begin{figure}[h]
\includegraphics[width=1\linewidth]{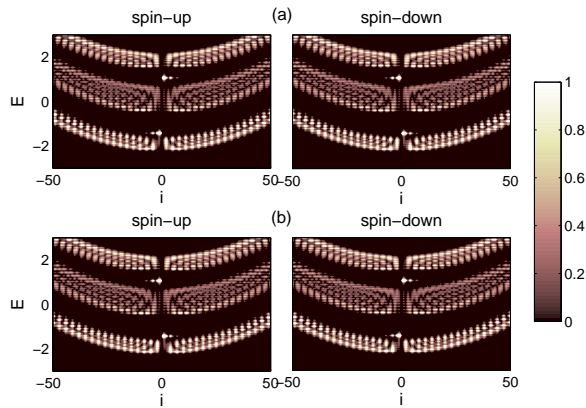}\newline
\caption{LDOS for spin-up and spin-down fermions with a single impurity of a large attractive potential ($F=-10$) within a slowly varying harmonic trap ($V_h=0.0005$). The midgap states for opposite spins in the bulk gaps are exchanged from one side of the impurity to the other by tuning the phase shift, (a) $\delta=\pi/3$, (b) $\delta=\pi$. Here we take $\alpha=1/3, t=1, L=201, V=1$.}
\label{mLDOS}
\end{figure}

Within the same harmonic trap independent of spins in cold atom experiments as the spinless one, we can also measure the LDOS for each spin using the SRRFS. As the relative phase $\delta$ changes, the positions of spin-up and spin-down midgap states will be exchanged (Fig. \ref{mLDOS}).

\section{Photonic crystals and the adiabatic pumping} \label{pump}
Due to the realization of the superlattices in photonic crystals
\cite{YEKraus2012}, another way to realize and detect the topology
of superlattices is with the help of photonic crystals. In the
experiment of Ref. \cite{YEKraus2012}, the superlattice can be
implemented by a set of coupled single-mode waveguides and the
propagating light can hop between the neighboring waveguides. The
on-site and the hopping terms of the superlattice can be
controlled by the refraction index of the waveguides and the
spacing between them, respectively. The propagating direction can
be used to detect the evolution of the injected light. In this
setup, we do not need harmonic trap and it is more easily to add a
single on-site impurity by just tuning one waveguide's refraction
index. So the above phenomena for the midgap states can be also
seen in this photonic crystal setup. If we inject a light into the
waveguide nearest to the impurity waveguide, we will see a
localized or an extended light signal coming out depending on the
$\delta$ we choose for the superlattice. To realize the spinful
superlattices, two kinds of injected light to feel the difference
of the lattice potential are needed.

\begin{figure}[h]
\includegraphics[width=0.9\linewidth]{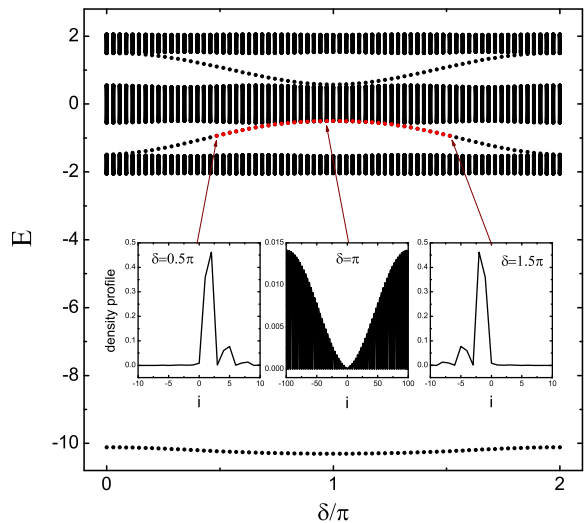}\newline
\caption{The eigenenergies versus $\delta$ for the 'off-diagonal' version of superlattices (\ref{offHam}) with a single on-site impurity of a large ($F=-10$) on-site potentials under the periodic modulation, $\alpha=1/3$. The pumping process of the midgap state are painted red. The insets are the density profiles show the initial ($\delta=0.5\pi$), middle ($\delta=\pi$), and final ($\delta=1.5\pi$) points of the $\delta$ to demonstrate the transfer from one side of the impurity to the other. The other parameters we choose are $t=1, L=201, V=1$ under PBCs.}
\label{offdiag}
\end{figure}

Besides the detection of the midgap states, we can also use the
photonic crystals to realize the adiabatic pumping of the midgap
states from one side of the impurity to the other. This process
should rely on the 'off-diagonal' version of the on-site modulated
superlattices, which has the similar properties as the on-site one
and can be realized by varying the spacings of waveguides along
the propagating direction. The Hamiltonian is as follows:
\begin{equation}
H=-\sum_{i}[(t+t_{i})\hat{c}_{i}^{\dag
}\hat{c}_{i+1}+\text{H.c.}]+F\hat{n}_{0},\label{offHam}
\end{equation}
where $t_{i}=V\cos (2\pi \alpha i+\delta )$ and $t$ is set to the
unit of energy. To realize the adiabatic pumping in this
``off-diagonal" version of superlattices, the propagating axis is
used to mimic the slowly varying of the phase shift $\delta$ of
the whole superlattice by changing the spacings of the waveguides
along this axis. As the injected light propagates in this set of
waveguides, the $\delta$ varies as well \cite{YEKraus2012}. Fig.
\ref{offdiag} shows the spectrum of the Hamiltonian (\ref{offHam})
with respect to $\delta$. If we inject a light into the waveguides
right-nearest to the impurity one at $\delta=0.5\pi$ in the
beginning, it will be localized at this site, and then as the
light propagates, the $\delta$ will be changed and the light will
be expanded to other sites at $\delta=\pi$, and finally it will
become localized at the left side of the impurity at
$\delta=1.5\pi$. So through this process we can pump the midgap
states from one side of the impurity to the other. Likewise, for
spin-1/2 fermions we can also pump the spins from one side to the
other or exchange different spins. This kind of manipulation may
be used in quantum information technique to control the states or
spins.

\section{Summary} \label{sum}
In summary, we show that a single on-site impurity will induce midgap states in the 1D topologically nontrivial superlattices. In the limit of a large impurity potential, these states will be localized next to the impurity site and behave like the edge states under OBCs. Based on this character of the midgap states, we propose an easier and more realistic scheme to detect the topology of 1D superlattices. In cold atoms, we calculate the LDOS for the SRRFS measurement to demonstrate the localization and the transfer of the midgap states for the spinless and spin-1/2 1D superlattices. Alternatively, the photonic crystal scheme is also proposed to demonstrate the feature of the midgap states and can be used to realize the adiabatic pumping. Our proposals are supplying a good way to show the topology of 1D superlattices and may be also helpful to the manipulation of the midgap states for quantum information.

\begin{acknowledgments}
This work has been supported by National Program for Basic
Research of MOST, NSF of China under Grants No.11121063 and
No.11174360, and 973 grant.
\end{acknowledgments}

\end{document}